\documentclass{article}
\usepackage{style}
\usepackage{shortcuts}

\usepackage{booktabs}
\usepackage{tabularx}

\title{Faster Approximate Text-to-Pattern Hamming Distances\\via Near-Optimal Sumset Estimation}
\author{%
    Nick Fischer \\[.3ex]\parbox[t]{0.25\textwidth}{\centering\normalsize Max Planck Institute for Informatics} \and 
    Yael Kirkpatrick\\[.3ex]\parbox[t]{0.25\textwidth}{\centering\normalsize MIT} \and
    Vasileios Nakos\\[.3ex]\parbox[t]{0.25\textwidth}{\centering\normalsize National and Kapodistrian University of Athens}}
\date{}

\begin{document}

\maketitle

\begin{abstract}
\noindent
\emph{Text-to-Pattern Hamming Distances} is one of the most studied problems in string algorithms: given a text $T$ of length~$n$ and a pattern~$P$, the task is to compute the Hamming distance between~$P$ and every length-$|P|$ substring of~$T$. As exact algorithms have been stuck at time~$\tilde O(n^{3/2})$ for decades, a long line of work has turned to the $(1+\epsilon)$-approximate version of the problem. Milestones include Karloff's $\tilde O(\epsilon^{-2} n)$-time algorithm [IPL '93], followed by randomized improvements to $\tilde O(\epsilon^{-1} n)$ by Kopelowitz and Porat [FOCS '15] and recently to $\tilde O(\epsilon^{-0.915} n)$ by Chan, Jin, Vassilevska W., and Xu [FOCS '23]. 

In this work, we significantly improve this bound to $\tilde O(\epsilon^{-2/3} n)$, using a considerably simpler algorithm. Our algorithm follows the framework of Chan, Jin, Vassilevska W., and Xu, which reduces the problem to a variant of 3SUM that we call \emph{Sumset Estimation}. We solve Sumset Estimation optimally (up to logarithmic factors) by replacing their matrix multiplication techniques with ideas from sparse recovery. Specifically, we adapt the classic Count Sketch by replacing its random signs with random complex roots of unity---a simple idea that may find further applications for other additive problems.

Finally, we give evidence that our running time might in fact be optimal: we show that the $\epsilon^{-2/3}$ dependence is \emph{best possible} for a certain natural class of algorithms, including ours and all previous algorithms.
\end{abstract}

\thispagestyle{empty}
\setcounter{page}{0}
\clearpage

\tableofcontents
\thispagestyle{empty}
\setcounter{page}{0}
\clearpage

\section{Introduction}
\emph{Text-to-Pattern Hamming Distances} (also known as \emph{Sliding-Window Hamming Distances}) is one of the most fundamental and best-studied problems in string algorithms. Given a text $T$ of length $n$ and a pattern $P$ of length $m \leq n$ over some common alphabet $\Sigma$, the task is to compute the Hamming distance (i.e., the number of mismatches) between $P$ and every length-$m$ substring of $T$, that is, to compute
\begin{equation*}
    \HD(T\intervalcc{i}{i + m - 1}, P)
\end{equation*}
for all shifts $i = 0, \dots, n - m$.

\paragraph{Exact Algorithms}
If the text is much longer than the pattern, a standard trick is to split the text into $O(n/m)$ overlapping pieces of length $2m$ and to solve each piece independently. For this reason, we assume that $n = \Theta(m)$ throughout and state all bounds in terms of $n$. The first nontrivial algorithm for Text-to-Pattern Hamming Distances is due to Fischer and Paterson~\cite{FischerP74}, who computed one convolution per alphabet symbol using the Fast Fourier Transform (FFT), leading to a running time of $O(|\Sigma| n \log n)$. This is near-linear for small alphabets, but becomes slow as the alphabet grows. For general alphabets, Abrahamson's classic algorithm~\cite{Abrahamson87} runs in time $O(n^{3/2} \sqrt{\log n})$. Remarkably, this algorithm remained unbeaten for more than three decades, until Chan, Jin, Vassilevska Williams, and Xu~\cite{Chan0WX23} recently shaved the remaining logarithmic factor and obtained an $O(n^{3/2})$-time algorithm. See also~\cite{JinX24} for a deterministic $O(n^{3/2} \sqrt{\log \log n})$-time algorithm. A long line of research has also studied the \emph{$k$-bounded} version of the problem (also known as the \emph{$k$-mismatch} problem), in which only the Hamming distances up to a given threshold~$k$ have to be reported~\cite{LandauV86,GalilG86,LandauV89,SahinalpV96,ColeH02,AmirLP04,CliffordFPSS16,GawrychowskiU18,ChanGKKP20,Charalampopoulos20,Chan0WX23}, culminating in an $O(n + k \sqrt{n})$-time algorithm~\cite{Chan0WX23}. For the general problem, however, it remains a major challenge to improve the exponent $3/2$.

\paragraph{Approximation Algorithms}
This $n^{3/2}$ barrier has motivated a parallel line of research on the \emph{$(1 + \epsilon)$-approximate} Text-to-Pattern Hamming Distances problem, where the goal is to approximate each Hamming distance up to a multiplicative factor of $1 + \epsilon$. This problem was first studied by Karloff~\cite{Karloff93}, who designed a randomized algorithm based on random projections that runs in time $O(\epsilon^{-2} n \log^2 n)$. Since then, a major challenge has been to optimize the dependence on $\epsilon$; see \cref{tab:previous} for an overview. Karloff's $\epsilon^{-2}$ dependence remained the state of the art for more than two decades (see also~\cite{Indyk98a} for an alternative, sampling-based algorithm), and it was even believed to be unavoidable: it stems from the variance of random projections, and it matches the $\Omega(\epsilon^{-2})$ lower bound for estimating the Hamming distance in the one-way communication model~\cite{Woodruff04,JayramKS08}. It thus came as a surprise when Kopelowitz and Porat~\cite{KopelowitzP15} broke this barrier with an $\tilde O(\epsilon^{-1} n)$-time\footnote{Throughout, $\tilde O(T) = T(\log T)^{O(1)}$ suppresses polylogarithmic factors.} algorithm based on ideas from sparse recovery. They later obtained a simplified algorithm avoiding the sparse recovery techniques~\cite{KopelowitzP18}. Chan, Golan, Kociumaka, Kopelowitz, and Porat~\cite{ChanGKKP20} showed that the problem can be solved in near-linear time $\tilde O(n)$ if $\epsilon$ is not too small (roughly, if $n \geq \epsilon^{-28}$). For smaller $\epsilon$, their approach inherently cannot beat $\tilde O(\epsilon^{-1} n)$.

However, small values of $\epsilon$ are the most interesting ones. For instance, for $\epsilon = n^{-1/2}$ the bound $\tilde O(\epsilon^{-1} n)$ coincides with the $O(n^{3/2})$ running time of the exact algorithms, so approximation does not seem to help at all. In a recent breakthrough, Chan, Jin, Vassilevska Williams, and Xu~\cite{Chan0WX23} achieved the first \emph{sublinear} dependence on $\epsilon^{-1}$, with an algorithm running in time $\tilde O(\epsilon^{-8/(11 - \omega)} n)$, where $\omega$ is the exponent of matrix multiplication. With the recent bound $\omega \leq 2.25$~\cite{OpenAI26} this becomes $\tilde O(\epsilon^{-0.915} n)$,\footnote{The running time stated in~\cite{Chan0WX23} is $\tilde O(\epsilon^{-0.928} n)$, which is based on the bound on $\omega$ known at the time.} and even if $\omega = 2$, it would be $\tilde O(\epsilon^{-8/9} n)$. Their algorithm is built on very different and rather intricate techniques from fine-grained complexity~\cite{ChanWX23}, combined with fast matrix multiplication.

Most of these algorithms are randomized, and for a long time the fastest deterministic algorithm remained Karloff's algorithm, which can be derandomized at the cost of one more log-factor in time $\tilde O(\epsilon^{-2} n)$. Recently, Fischer, Jin, and Xu~\cite{Fischer0X25} improved the deterministic dependence to $\epsilon^{-1}$, with a running time of $\epsilon^{-1} n^{1+o(1)}$. In terms of deterministic algorithms, the $\epsilon^{-1}$-dependence hits a natural barrier. For further background on Text-to-Pattern Hamming Distances and related problems (such as approximating other distance measures), we refer to the survey by Uzna\'{n}ski~\cite{Uznanski20}.

\subsection{Our Improvement}
In this paper, we improve the dependence on $\epsilon$ to $\epsilon^{-2/3}$:

\begin{theorem} \label{thm:main}
    The $(1+\epsilon)$-approximate Text-to-Pattern Hamming Distances problem can be solved in time $\tilde{O}(\epsilon^{-2/3} n)$ by a randomized algorithm.
\end{theorem}

This improves upon the previous best $\tilde O(\epsilon^{-0.915} n)$-time algorithm~\cite{Chan0WX23}, and it even beats the bound $\tilde O(\epsilon^{-8/9} n)$ that their algorithm would achieve if $\omega = 2$. For instance, in the interesting regime $\epsilon = n^{-1/2}$ our algorithm runs in time $\tilde O(n^{4/3})$, whereas the previous best bound was $\tilde O(n^{1.458})$. Another benefit is that our algorithm is considerably simpler: it does not rely on fast matrix multiplication at all, and at its core it uses little more than random sampling and FFT.

\begin{table}[t]
\renewcommand{\arraystretch}{1.5}
\setlength{\tabcolsep}{5pt}
\begin{tabularx}{\textwidth}{@{}>{\raggedright}p{.41\textwidth}@{}p{.22\textwidth}@{}p{.35\textwidth}<{\raggedright}@{}}
\toprule
\textbf{Reference} & \textbf{Running Time} & \textbf{Techniques} \\
\midrule
    Karloff~\cite{Karloff93} & $\tilde O(\epsilon^{-2} n)$\,$^{\ast}$ & random projection \\
    Indyk~\cite{Indyk98a} & $\tilde O(\epsilon^{-3} n)$ & random sampling \\
    Kopelowitz and Porat~\cite{KopelowitzP15} & $\tilde O(\epsilon^{-1} n)$ & random projection + sparse recovery \\
    Kopelowitz and Porat~\cite{KopelowitzP18} & $\tilde O(\epsilon^{-1} n)$ & random projection \\
    Chan, Golan, Kociumaka, Kopelowitz, Porat~\cite{ChanGKKP20} & $\tilde O(n)$ for $n \geq \epsilon^{-28}$ & random sampling + rect.\ matrix mult. \\
    Chan, Jin, Vassilevska Williams, Xu~\cite{Chan0WX23} & $\tilde O(\epsilon^{-0.915} n)$ & \#3SUM + ``Fredman's trick meets dominance product'' \\
    Fischer, Jin, Xu~\cite{Fischer0X25} & $O(\epsilon^{-1} n^{1+o(1)})$\,$^{\ast}$ & \#3SUM, deterministic approximate  \\
    \textbf{This paper} & $\tilde O(\epsilon^{-2/3} n)$ & \#3SUM + sparse recovery \\
\bottomrule
\end{tabularx}
\caption{Approximation algorithms for Text-to-Pattern Hamming Distances, focusing on the dependence on~$\epsilon$ and ignoring logarithmic factors (assuming that $n = \Theta(m)$). Algorithms marked with~$^\ast$ are deterministic.} \label{tab:previous}
\end{table}

Our algorithm builds on the framework of~\cite{Chan0WX23}, which reduces approximate Text-to-Pattern Hamming Distances to the following variant of 3SUM that we call \emph{Sumset Estimation}: Given two sets of integers $A, B$ and a parameter $\sigma$, compute unbiased estimates with variance at most $\sigma^2$ of the number of representations $r_{A+B}(z) = \abs{\set{(a, b) \in A \times B : a + b = z}}$ of every sum $z$. Our main technical contribution is a simple, near-optimal algorithm for this problem, running in time $\tilde O(|A| + |B| + |A| \, |B| / \sigma)$ (\cref{thm:sumset-estimate}); this matches the worst-case output size and is thus optimal up to logarithmic factors. In contrast to~\cite{Chan0WX23}, we avoid Fredman's trick and fast matrix multiplication altogether, and instead rely on techniques from \emph{sparse recovery}. Specifically, we adapt the classic Count Sketch~\cite{CharikarCF02} by replacing its random signs with random complex roots of unity. This seemingly small change makes the sketch compatible with convolutions, so that it can be computed efficiently by FFT. We believe that this adapted Count Sketch is an interesting tool in itself that may find further applications to additive problems. Along the way, we also simplify parts of the reduction from~\cite{Chan0WX23}. We explain all of these ideas in more detail in our technical overview.

\paragraph{Is This Optimal?}
In a sense, our algorithm pushes the approach of~\cite{Chan0WX23} to its limit: we leave their reduction to Sumset Estimation essentially unchanged, and our algorithm for Sumset Estimation is optimal up to logarithmic factors, so there is no room for further improvement on this front. But could a different approach lead to an even better dependence on~$\epsilon$? In \cref{sec:hard-instance} we give some evidence that this might be difficult: We consider a natural class of algorithms that independently estimate the number of matches involving each symbol. In particular, our new algorithm as well as the predecessor by Chan, Jin, Vassilevska~W., and Xu~\cite{Chan0WX23} are part of this class. We show that, within this class, our algorithm is essentially the fastest possible. The critical regime turns out to be $\epsilon = n^{-3/4}$. Here our algorithm runs in time $\tilde O(n^{3/2})$, which matches the exact algorithms, and our hard instance suggests that this is no coincidence. In this regime, beating time $n^{3/2}$ seems to require techniques that would also lead to a breakthrough for the exact problem. We stress, however, that this barrier only applies to a restricted class of algorithms, and it is entirely possible that it can be overcome by fundamentally different techniques.
\section{Preliminaries}
A string $S$ of length $s = |S|$ is a sequence of $s$ characters $S[0]S[1]\ldots S[s-1]$ over an alphabet $\Sigma$. For indices $0\leq i\leq j\leq s-1$ denote the substring $S[i]S[i+1]\ldots S[j]$ by $S\intervalcc{i}{j}$. A run of a string is a maximal consecutive block of identical characters. The run-length encoding (RLE) of a string is the number of runs in it.

The Hamming distance of two strings is the number of indices at which the strings do not match,
\[\HD(S_1,S_2) = |\{i: S_1[i]\neq S_2[i]\}|.\]

For integers $N, M$, write $[N, M] = \set{N, \dots, M}$ and $[N] = \set{0, \dots, N-1}$. For two sets of integers $A, B$, define the \emph{sumset} $A + B = \set{a + b : a \in A,\, b \in B}$ and denote the \emph{multiplicity} of an element $z \in A + B$ by $r_{A+B}(z) = \abs{\set{(a,b) \in A \times B : a + b = z}}$. 

For two functions $f, g : [0, N] \to \Complex$, define the convolution $f * g : [0, 2N] \to \Complex$ by
\begin{equation*}
    (f * g)[z] = \sum_{\substack{x, y \in [0, N]\\x + y = z}} f(x) \cdot g(y).
\end{equation*}
Similarly, define the \emph{cyclic} convolution where the condition $x + y = z$ is replaced by $x + y \equiv z \mod N$. It is well-known that both the convolution and the cyclic of two functions can be computed in time $O(N \log N)$ by the Fast Fourier Transform.

Define a multiplicative $(1+\epsilon)$-approximation to a function $f$ as a function $\tilde{f}$ on the same domain such that for any input $x$ in the domain $(1-\epsilon)f(x) \leq \tilde{f}(x)\leq (1+\epsilon) f(x)$. We define an additive $\beta$-approximation similarly, with the requirement that $f(x) - \beta \leq \tilde{f}(x)\leq f(x) + \beta$.

\section{Technical Overview}
Before we give the detailed proof, let us start with an informal overview of the ideas behind our algorithm.

\subsection{The Chan--Jin--Vassilevska W.--Xu Algorithm}
Our algorithm works in the same framework as the previous algorithm due to Chan, Jin, Vassilevska W., and Xu~\cite{Chan0WX23}. To set the stage, we will therefore first describe the key ideas behind that algorithm. It is designed around the following computational problem:\footnote{In~\cite{Chan0WX23}, the authors refer to this problem instead as \emph{Approximate Counting All-Numbers 3SUM}, but we prefer to stick to the shorter name \emph{Sumset Estimation}.}

\begin{definition}[Sumset Estimation]
Given sets $A, B \subseteq [0, N]$ and a parameter $\sigma \geq 1$, compute a sparsely represented function $r : [0, 2N] \to \Real$ such that
\begin{enumerate}[label=(\arabic*)]
    \item $\Ex[r(z)] = r_{A+B}(z)$ for all $z \in [0, 2N]$, and
    \item $\Var[r(z)] \leq \sigma^2$.
\end{enumerate}
\end{definition}

The algorithm can then be seen as an elaborate reduction from Text-to-Pattern Hamming Distances to Sumset Estimation. These two steps can be qualitatively summarized as follows; quantitatively, it is somewhat messy to trace the choice of parameters throughout, so we will omit them.

\paragraph{Step 1: Reduction to Sumset Estimation}
The reduction can be split into three substeps. We will mostly leave these steps untouched, but they are helpful to understand how Sumset Estimation is related to pattern matching:

\begin{enumerate}
    \item \emph{From Multiplicative to Additive Approximation:} The goal is a multiplicative $(1 + \epsilon)$-approximation for Text-to-Pattern Hamming Distances. Based on previous work by Gawrychowski and Uzna\'{n}ski~\cite{GawrychowskiU18}, this can be reduced to computing an \emph{additive} $\epsilon k$-approximation for the case when both strings have \emph{run-length encodings} of size at most $k$. Additive approximations are more convenient, notably since it is equivalent to compute additive approximations of the number of \emph{mismatches} and the number of \emph{matches}.
    \item \emph{From Hamming Distances to Estimating Convolutions:} Next, we reduce to a variant of Sumset Estimation. Specifically, enumerate all characters $s$ that appear in text $T$ and pattern~$P$. Let $A_s, B_s$ denote the sets of positions where $s$ appears in text and in the \emph{reversed} pattern. Then the number of matches at shift $i$ is exactly
\begin{equation*}
    \sum_s \abs{\set{j : T[j + i] = P[j] = s}} = \sum_s r_{A_s + B_s}(|P| - 1 + i).
\end{equation*}
In other words, the quantity we want to estimate can be expressed as the sum of many sumset multiplicities, one for each character $s$. To compute this estimate, apply a simple heavy/light trick. For symbols appearing frequently, compute $r_{A_s + B_s}$ exactly by one expensive FFT. For infrequent symbols, instead approximate $r_{A_s + B_s}$ by Sumset Estimation. This is faster, but leads to some estimation error. This basic heavy/light trick dates back to the classic algorithm for (exact) Text-to-Pattern Hamming Distances due to Abrahamson~\cite{Abrahamson87}; the difference is that in Abrahamson's algorithm the infrequent symbols are handled by brute force.
    \item \emph{From Run-Length Encoded to Sparse Functions:} In substep 1 we have reduced to run-length encoded strings, and so naturally the resulting sets $A_s, B_s$ in substep~2 also have small run-length encodings, i.e., they can be described as the union of few intervals. However, they are not necessarily sparse. For this reason, this third reduction step reduces run-length encoded sets to truly sparse sets. It remains to solve multiple instances of the Sumset Estimation problem defined above, one for each infrequent character.
\end{enumerate}

\paragraph{Step 2: Solve Sumset Estimation} In the second step they efficiently solve Sumset Estimation by an algorithm in time $\tilde O(n^{(3+\omega)/2} / \sigma^{(1+\omega)/2})$ for the case when \makebox{$|A|, |B| \leq n$} (and when~\makebox{$\sigma \geq \sqrt{n}$}) \cite[Lemma~3.3]{Chan0WX23}. This is achieved by another heavy/light trick: the small multiplicities $r_{A+B}(z)$ can be estimated by naive sampling, and only the large multiplicities require more care. These are estimated in~\cite{Chan0WX23} based on the powerful (yet quantitatively lossy) ``Fredman's trick meets dominance product'' technique pioneered in~\cite{ChanWX23} and fast matrix multiplication. Strictly speaking, this heavy/light idea is not implemented directly for Sumset Estimation, but instead by yet another reduction, namely to a variant of the \emph{Exact Triangle} problem.\footnote{In this problem the input consists of an edge-weighted tripartite graph. The goal is to estimate, for each edge, in how many triangles of zero total weight the edge participates. As for the convolution problem, the goal is to achieve an unbiased estimator with small variance.}

\subsection{Our Improvement: A Near-Optimal Algorithm for Sumset Estimation}
In our algorithm we leave Step 1 almost untouched; we only improve it in terms of some technical aspects leading to a cleaner and simpler overall reduction. Our main contribution is a new algorithm for Step 2. We avoid the ``Fredman's trick meets dominance product'' technique and fast matrix multiplication altogether. Based on very different techniques from sparse recovery, we obtain a \emph{near-optimal} algorithm for Sumset Estimation: 

\begin{theorem}[Near-Optimal Sumset Estimation] \label{thm:sumset-estimate}
There is a randomized algorithm that, given \makebox{$A, B \subseteq [0, N]$} and parameters $\sigma, \rho > 0$, computes estimates $r$ such that:
\begin{enumerate}[label=(\arabic*)]
    \item $\Ex[r(z)] = r_{A+B}(z) \pm \rho$,
    \item $\Var[r(z)] \leq \sigma^2$, and
    \item the expected number of nonzero estimates $r(z)$ is at most $O(|A| \, |B| / \sigma)$.
\end{enumerate}
The algorithm runs in expected time \smash{$(|A| + |B| + |A| \, |B| / \sigma) \cdot \log^{O(1)} (N / \rho)$}.
\end{theorem}

To digest this statement, first note that it is crucial that the estimates $r$ are returned \emph{sparsely}, i.e., as a list of the nonzero estimates. If the algorithm were to return an estimate for each $z \in A + B$, the output size could already reach $|A| \, |B|$. But even if the estimates are returned sparsely, the output size is $\Omega(|A| \, |B| / \sigma)$ in the worst case.\footnote{To see this, choose any $1 \leq \sigma \leq n$ and let $N = n^2 / \sigma$. Pick $A = \set{1, \dots, n}$ and pick $B = \set{\floor{N/n}, 2\floor{N/n}, \dots, n \floor{N / n}}$. Both sets have size $|A|, |B| = \Theta(n)$, and their sumset has size $\Omega(N)$. Moreover, a constant fraction of the sumset entries $z$ has multiplicity $r_{A+B}(z) \geq \Omega(n^2 / N) \geq \Omega(\sigma)$. Each such entry satisfies $\Ex[r(z)]^2 \geq \Omega(\sigma^2)$, and therefore the Paley--Zygmund inequality implies that
\begin{equation*}
    \Pr(r(z) > 0) \geq \frac{\Ex[r(z)]^2}{\Ex[r(z)^2]} = \frac{\Ex[r(z)]^2}{\Var[r(z)] + \Ex[r(z)]^2} = \Omega(1).
\end{equation*}
(This inequality applies even though $r(z)$ may be negative since $\Ex[r(z)] > 0$.)
In particular, we expect $r(z)$ to be nonzero for a constant fraction of the sumset entries, and thus the expected output size is $\Omega(N) = \Omega(n^2 / \sigma)$.} For this reason, \cref{thm:sumset-estimate} is \emph{unconditionally optimal}, up to log-factors.

For technical reasons, \cref{thm:sumset-estimate} suffers from a small precision error $\rho$; i.e., the returned estimates are biased slightly by $\pm \rho$. For all reasonable purposes, we can set $\rho = N^{-100}$, which leads to negligible error overall, and which does not worsen the running time asymptotically. It will become clear later why our approach naturally leads to some small precision error $\rho$.

\paragraph{Proof Ideas behind \cref{thm:sumset-estimate}}
As in \cite{Chan0WX23}, we use the idea that all sumset entries with small multiplicity, namely with $r_{A+B}(z) \leq \sigma$, can be estimated with appropriate precision by a naive sampling algorithm. Specifically, sample elements $a_1, \dots, a_q \in A, b_1, \dots, b_q \in B$ uniformly and independently at random. Then estimate the multiplicity $r_{A+B}(z)$ by extrapolating how many samples satisfy $z = a_i + b_i$. Choosing~\makebox{$q = \Theta(n^2 / \sigma)$} (matching our time budget) leads to estimates with variance $r_{A+B}(z) \cdot \sigma$. Whenever $r_{A+B}(z) \leq \sigma$, this is sufficiently accurate.

The hard part is to estimate the multiplicities larger than $\sigma$. For this overview let us assume that we know the set $H = \set{z \in A + B : r_{A+B}(z) \geq \sigma}$; in the actual algorithm, we will instead use another sampling algorithm to approximate $H$. In the language of sparse recovery, $H$ can be viewed as the \emph{heavy hitters} of the function $r_{A+B}(z)$, and our task of estimating $r_{A+B}(z)$ with bounded variance can be viewed as \emph{point queries} to $f := r_{A+B}$. This problem is well-studied in the sparse recovery, sketching, and streaming literature: we access the vector $f$ via linear queries (i.e., inner products with~$f$), and the goal is to obtain accurate estimates using few queries. It has a prominent solution: the \emph{Count Sketch}~\cite{CharikarCF02}. Our proof for \cref{thm:sumset-estimate} is essentially a Count Sketch tailored for convolutions, and for this reason we first take a quick excursus into sparse recovery.

\paragraph{Excursus: Count Sketch}
Fix some function $f : [N] \to \Real$, which can be equivalently viewed as a vector $f \in \Real^N$. In a typical sparse recovery setup, we think of $f$ as \emph{approximately} sparse, i.e., as dominated by a few large entries plus noisy small entries. The general goal is to query $f$ nonadaptively to subsequently answer queries about $f$, e.g., to approximate the interesting entries of $f$. These nonadaptive queries to $f$ are typically \emph{linear queries}; in such a query we specify a query vector $q \in \Real^N$ and the output is $q^T f = \sum_x q(x) f(x)$.

The \emph{Count Sketch} due to Charikar, Chen, and Farach-Colton~\cite{CharikarCF02} is a core primitive in sparse recovery. It is a randomized algorithm (or data structure if you wish) that, for any parameter $k$, prepares \smash{$ O(k)$} linear queries such that, when provided answers to these linear queries for $f$, we can subsequently estimate arbitrary entries of $f$ with good precision. Specifically, we can query the sketch with some $x$ and it returns an unbiased estimate $r(x)$ of $f(x)$ with variance $\Var[r(x)] \leq \norm{f}_2^2 / k$.

The construction behind Count Sketch is simple and elegant. It constructs $p:=O(k)$ queries using a hash function $h:[N] \rightarrow [p]$, so that every coordinate~$x$ participates in exactly one of these queries, chosen at random. Furthermore, it chooses a random sign for each coordinate, i.e.,
$s : [N] \to \set{-1, 1}$. Then, it computes and stores for each $b \in [p]$ the quantity
\begin{equation*}
    Q_b := \sum_{x: h(x) = b} s(x) f(x).
\end{equation*}
A point query at $x$ can then be answered by returning  $r(x) := s(x) Q_{h(x)}$.

To see why this makes sense, put $i=h(x)$ and observe that the expectation and variance of the resulting estimates can be bounded by
\begin{equation*}
   \Ex[r(x)] = \Ex\brackets*{s(x) \cdot \sum_{y \in h^{-1}(i)} s(y) f(y)} = f(x) + \Ex\brackets*{\sum_{y \in h^{-1}(i) \setminus \set{x}} s(x) s(y) f(y)} = f(x),
\end{equation*}
and
\begin{align*}
   \Var[r(x)] &= \Ex\brackets*{\parens*{\sum_{y \in h^{-1}(i)} s(y) f(y)}^2} - f(x)^2 \\
   &= \Ex\brackets*{\sum_{y_1, y_2 \in h^{-1}(i) \setminus \set{x}} s(y_1) s(y_2) f(y_1) f(y_2)} \\
   &= \Ex\brackets*{\sum_{y \in h^{-1}(i) \setminus \set{x}} f(y)^2}.
\end{align*}
Here we have only used the randomness of the random signs, but the randomness of $h$ is still unresolved. Thus,
\begin{equation*}
   \Var[r(x)] = \frac{1}{k} \cdot \sum_{y \neq x} f(y)^2 \leq \frac{\norm{f}_2^2}{k}.
\end{equation*}

\paragraph{Sumset Estimation via an Adapted Count Sketch}
Let us finally return to our setting, where the goal is to estimate the entries of $f = r_{A+B}$ at all query points in $H$. We will apply the Count Sketch with $k = n^2 / \sigma$, dictated by our running time budget. There are two issues, both requiring adaptations to the sketch.

The first problem is that the variance bound $\Var[r(x)] \leq \norm{f}_2^2 / k$ is not good enough; $\norm{f}_2^2$ can be as large as $n^3$ and thus the variance is bounded only by $\sigma n$. This is a technical issue with a simple fix: for each query point~$x$ we will \emph{make sure} to select a hash function satisfying

\begin{equation} \label{sec:overview:eq:l1}
    \sum_{y \neq x : h(y) = h(x)} f(y) \leq \frac{\norm{f}_1}{k} = \frac{|A| \, |B|}{k};
\end{equation}
a randomly chosen $h$ still satisfies this property with constant probability. Then the variance, depending only on the randomness of the signs $s$, can indeed be bounded by
\begin{equation*} 
   \Var[r(x)] \leq \sum_{y\neq x : h(y) = h(x)} f(y)^2 \leq \parens*{\sum_{y\neq x: h(y)=h(x)} |f(y)|}^2 \leq \frac{\norm{f}_1^2}{k^2} \leq \sigma^2.
\end{equation*}

The second issue is much more severe, and its resolution can be seen as the key insight behind our algorithm. Count Sketch requires access to arbitrary linear queries of $f$. However, recall that our goal is to compute $f$ to begin with, so we certainly cannot efficiently support arbitrary queries. Our approach is to design appropriate replacements of the queries from above, and the fact that $h$ and $s$ were both from universal hash families will come in handy.

In particular, we shall make $p$ slightly larger, of the order $O(k \log N)$, and choose it to be a random prime number in an interval of size $\Omega(k \log N)$, then set $h(x) = x~\mathrm{mod}~p$. This ensures, by the Prime Number Theorem, that two distinct elements collide under~$h$ with probability at most $O(1/k)$. We can then access all queries of the form
\begin{equation}
    Q_b  =\sum_{x: h(x)=b} f(x),
\end{equation}
for all $b \in [p]$, in time $O(p \log p + |A| + |B|)$. Indeed, let $g, g' : \Int / p\Int \to \Int_{\geq 0}$ be the functions counting how many elements of~$A, B$ fall into the residue classes modulo $p$, i.e., $g(i) = \abs{\set{a \in A : a \equiv i \mod p}}$. Compute the cyclic convolution of $g$ and $g'$ via one FFT in time $ O(p \log p)$. Then $Q_b$ is exactly the $b$-th entry in the cyclic convolution. In other words, we can access the hashed version of the convolution in time $O(p \log p) = O(k \log^2 N)$ by convolving the hashed indicator vectors of $A$ and $B$.


Of course, the queries $Q_b$ do not account yet for the random signs. And a priori, it seems impossible to incorporate them. A first, wrong attempt could be to adapt $g$ to
\begin{equation*}
    g(i) = \sum_{\substack{a \in A\\h(a)=i}} s(a),
\end{equation*}
for some randomly chosen signs $s$. However, then the convolution $g * g'$ is useless, since two terms $a_1 + b_1 = a_2 + b_2 = z$ may contribute to $z \bmod p$ positively and negatively and may therefore cancel. Instead, we need a substitute $s$ for random signs that interplays with taking convolutions, i.e., which satisfies $s(a) s(b) = s(a + b)$ and retains the cancellations in the previous calculations. Our solution is to work with \emph{random complex roots of unity}. The idea of working with roots of unity, or more generally with elements of the complex numbers or of finite fields, has appeared before in~\cite{nakos2020nearly, bringmann2021sparse}, but for a different purpose, mostly to locate the support of the convolution. Using roots of unity to simulate random signs in a pairwise independent manner is most closely related to ideas from sparse Fourier transforms and $F_2$ norm estimation in streaming algorithms~\cite{kane2011fast,indyk2014nearly}.

Recall that $\omega = \exp(2\pi i / N) \in \Complex$ is an \emph{$N$-th root of unity} since it satisfies $\omega^N = 1$. Moreover, all integer powers of $\omega$ are $N$-th roots of unity as well. Pick a random $\ell$, and replace the sign function $s : [N] \to \set{-1, 1}$ by $s : [N] \to \Complex$ defined by $s(x) = \omega^{\ell x}$. Then the Count Sketch analysis still works out (modulo appropriate technical changes), based on the following basic fact:

\begin{fact} \label{fact:random-root-of-unity}
Let $u \geq 1$ be an integer, let $\omega = \exp(2\pi i / u) \in  \Complex$, and let $0 \leq \ell < u$ be an integer drawn uniformly at random. Then for all $x \in \Int$:
\begin{equation*}
    \Ex[\omega^{\ell x}] =
    \begin{cases}
        1 &\text{if $x \equiv 0 \mod u$,} \\
        0 &\text{otherwise.}
    \end{cases}
\end{equation*}
\end{fact}

Moreover, this alternative choice satisfies $s(a) s(b) = s(a + b)$ and therefore interplays nicely with taking convolutions. Specifically, consider the functions

\begin{equation*}
    g(i) = \sum_{\substack{a \in A\\h(a)=i}} \omega^{\ell a} \quad\text{and}\quad g'(i) = \sum_{\substack{b \in B\\h(b)=i}} \omega^{\ell b}.
\end{equation*}
Their cyclic convolution modulo $p$ provides exactly the desired answers to the queries $Q$.

All in all, this alternative Count Sketch tailored for convolutions appears to be a powerful tool which we expect to find applications for additive problems beyond just Sumset Estimation. Of course, computing with complex numbers introduces technical issues (like the precision overhead $\rho$ in \cref{thm:sumset-estimate}) which we will discuss in the technical part (\cref{sec:algo:sec:heavy:sec:complex}).

\section{Near-Optimal Sumset Estimation}

\subsection{Estimating Heavy Hitters}
This section is dedicated to proving the following lemma:

\begin{lemma}[Estimating Sumset Heavy Hitters] \label{lem:heavy-hitters}
There is a randomized algorithm that, given sets $A, B \subseteq [0, N]$ and $H \subseteq [0, 2N]$, and parameters $\sigma, \rho > 0$, computes estimates $r$ such that, for all $z \in H$,
\begin{enumerate}[label=(\arabic*)]
    \item $\Ex[r(z)] = r_{A+B}(z) \pm \rho$, and
    \item $\Var[r(z)] \leq \sigma^2$.
\end{enumerate}
The algorithm runs in expected time $(|A| + |B| + |H| + |A| \, |B| / \sigma) \cdot \log^{O(1)} (N / \rho)$.
\end{lemma}

\begin{algorithm}[htbp]
\caption{The sumset heavy hitter estimator (\cref{lem:heavy-hitters}).} \label{alg:heavy-hitters}
\begin{algorithmic}[1]
\Input{Sets $A, B \subseteq [0, N]$ and $H \subseteq [0, 2N]$}
\Output{Estimates $r(z) \approx r_{A+B}(z)$ for all $z \in H$}

\medskip
\State Let $u = 2N + 1$, let $p_0 = c \log u \cdot |A| \, |B| / \sigma$ for some constant $c$, and let $\omega := \exp(2\pi i / u)$
\Repeat
    \State\label{alg:heavy-hitters:line:p}\ignorespaces Sample a uniformly random prime $p$ in the interval $[p_0, 2p_0]$
    \State\label{alg:heavy-hitters:line:ell}\ignorespaces Sample $0 \leq \ell < u$ uniformly at random
    \State\label{alg:heavy-hitters:line:gh}\ignorespaces Compute the functions $g, h : \Int / p\Int \to \Complex$ defined by
    \begin{align*}
        g(i) &= \sum_{\substack{x \in A\\x \equiv i \mod p}} \omega^{\ell x}, \\
        h(j) &= \sum_{\substack{y \in B\\y \equiv j \mod p}} \omega^{\ell y}.
    \end{align*}
    \State\label{alg:heavy-hitters:line:f}\ignorespaces Compute the cyclic convolution $f$ of $g$ and $h$
    \State\label{alg:heavy-hitters:line:good}\ignorespaces Run \cref{lem:good-approx} to compute a subset $G \subseteq H$ of $\sigma$-good elements
    \ForEach{$z \in G$}\label{alg:heavy-hitters:line:good-loop}
        \State\label{alg:heavy-hitters:line:assign}\ignorespaces Set $r(z) := \omega^{-\ell z} \cdot f(z \bmod p)$
    \EndForEach
\Until{$r(z)$ has been assigned for all $z \in H$}
\end{algorithmic}
\end{algorithm}

The algorithm behind this lemma is summarized in pseudocode in \cref{alg:heavy-hitters}. The problem is trivial unless $1 \leq \sigma \leq \min\set{|A|, |B|}$, so we will occasionally assume these bounds. We set $u = 2N + 1$, $p_0 = c \log u \cdot |A| \, |B| / \sigma$ for some absolute constant $c$ to be determined, and $\omega := \exp(2\pi i /u)$, i.e., a complex $u$-th root of unity. The algorithm runs in rounds, where in each round we will assign some of the unassigned values $r(z)$, but not necessarily all. Once all values are assigned, the algorithm stops. Each round works as follows. We sample a uniformly random prime~$p$ in the interval $[p_0, 2p_0]$. We also sample $0 \leq \ell < u$ uniformly at random. Construct the (densely represented) functions $g, h : (\Int/p\Int) \to \Complex$ defined by
\begin{align*}
    g(i) &= \sum_{\substack{x \in A\\x \equiv i \mod p}} \omega^{\ell x}, \\
    h(j) &= \sum_{\substack{y \in B\\y \equiv j \mod p}} \omega^{\ell y}.
\end{align*}
Compute their cyclic convolution $f$:
\begin{equation*}
    f(k) = \sum_{\substack{i, j \in \Int/p\Int\\i+j\equiv k \mod p}} g(i) \cdot h(j).
\end{equation*}
Now, we want to assign $r(z) := \omega^{-\ell z} \cdot f(z \bmod p)$ for all $z \in H$; however, we will do so only for a selected subset of \emph{good} elements of $H$. This subset is computed by a separate algorithm, stated in \cref{lem:good-approx}; why will become clear in the analysis. As is, there remains one issue: the estimates are complex numbers. For now, let us ignore this issue, and further assume that all computations with complex numbers are exact. We will remove this assumption in \cref{sec:algo:sec:heavy:sec:complex}. Note that different elements of $H$ can use the \emph{same} convolution entry $f(z \bmod p)$. Their estimates may therefore be correlated, which is harmless because \cref{lem:heavy-hitters} only bounds the expectation and variance of each individual estimate.

\subsubsection{Good Elements and Running Time} \label{sec:algo:sec:heavy:sec:analysis}
For the analysis, focus on a single round. We will resolve the randomness of $p$ and~$\ell$ in two stages. Focus on the first stage where we sample $p$. We classify the elements of $H$ as follows:

\begin{definition}[Good Elements] \label{def:good}
We call~$z \in H$ \emph{$\tau$-good} if
\begin{equation*}
    W(z) := \sum_{\substack{z' \in [0, 2N] \setminus \set{z}\\z' \equiv z \mod p}} r_{A+B}(z') \leq \tau.
\end{equation*}
\end{definition}

Intuitively, this definition captures exactly the elements of $H$ for which the algorithm can produce an accurate estimate, and for that reason, it only assigns $r(z)$ for good elements. Of course, this requires computing the set of good elements first:

\begin{lemma} \label{lem:good-approx}
There is a randomized algorithm to compute a subset $G \subseteq H$ containing all $\sigma/4$-good elements and no elements that are not $\sigma$-good. The algorithm runs in time $\tilde O(|H| + |A| \, |B| / \sigma \cdot \log (N/\rho))$ and succeeds with probability $1 - (N/\rho)^{-10}$.
\end{lemma}
\begin{proof}
We rely on sampling. Let $q = \ceil{|A| \, |B| / \sigma}$ and sample $a_1, \dots, a_q \in A$ and~\makebox{$b_1, \dots, b_q \in B$} uniformly and independently. Write $z_i = a_i + b_i$. \emph{Mark} all~\makebox{$z \in H$} for which there is a sample $z_i$ with $z \neq z_i$ and $z \equiv z_i \mod p$. We analyze the probability of marking $z$. Each sample $z_i$ leads to a mark of $z$ with probability exactly $W(z) / (|A| \, |B|)$. If $W(z) \geq \sigma$, then the probability of not marking~$z$ is at most
\begin{equation*}
    \vphantom{\Bigg(}\parens*{1 - \frac{\sigma}{|A| \, |B|}}^q \leq \exp\parens*{-\frac{q \sigma}{|A| \, |B|}} \leq \exp(-1) \leq 0.4.
\end{equation*}
Conversely, if $W(z) \leq \sigma/4$, then the probability of not marking $z$ is at least
\begin{equation*}
    \vphantom{\Bigg(}\parens*{1 - \frac{\sigma/4}{|A| \, |B|}}^q \geq \exp\parens*{-\frac{2q \sigma}{4|A| \, |B|}} \geq \exp\parens*{-\frac12} \geq 0.6,
\end{equation*}
using the estimate $1 - x \geq \exp(-2x)$ for all $0 \leq x \leq 1/2$ (a condition that can be assumed here without loss of generality, since \cref{lem:heavy-hitters} is trivial for $\sigma \geq |A|$).

With this in mind, repeat this process $\Theta(\log(N/\rho))$ times, and include in $G$ all elements from $H$ that are marked in the minority of repetitions. Chernoff's bound implies that this includes all $\sigma/4$-good elements, but no element which fails to be $\sigma$-good, with high probability $1 - (N / \rho)^{-10}$. The total time is $O(|H| + q \log (N / \rho))$, which is as claimed.
\end{proof}

Before we turn to the correctness, let us first analyze the running time of \cref{alg:heavy-hitters}. To this end, we show that every element in $H$ is $\sigma/4$-good with constant probability:

\begin{lemma} \label{lem:good}
Each $z \in H$ is $\sigma/4$-good with probability at least $1/2$ (over the random choice of $p$).
\end{lemma}
\begin{proof}
Observe that
\begin{align*}
    \Ex[W(z)] &= 
    \Ex\brackets*{\sum_{\substack{z' \neq z\\z' \equiv z \mod p}} r_{A+B}(z')} \\
    &= \sum_{z' \neq z} r_{A+B}(z') \cdot \Pr(z \equiv z' \mod p) \\
    &\leq \sum_{z' \neq z} r_{A+B}(z') \cdot \frac{\log_{p_0}(u)}{\Omega(p_0 / \log p_0)} \\
    &\leq |A| \, |B| \cdot O\parens*{\frac{\sigma}{c |A| \, |B|}} \\
    &\leq \frac{\sigma}{8},
\end{align*}
by the Prime Number Theorem and by choosing the constant $c$ sufficiently large. By Markov's inequality it follows that $W(z) \leq \sigma/4$ with probability at least $1/2$.
\end{proof}

We show that \cref{alg:heavy-hitters} runs in expected time $(|A| + |B| + |H| + |A| \, |B| / \sigma) \cdot \log^{O(1)} N$ as claimed. We first argue that the expected number of iterations is $O(\log N)$. Indeed, \cref{lem:good} implies that any fixed element $z \in H$ is $\sigma/4$-good with probability at least $1/2$. Any such element is included into $G$ with high probability by \cref{lem:good-approx}; once this happens we assign $r(z)$. Thus, after $O(\log N)$ rounds we expect to have assigned all estimates.

We now bound the running time of a single round. Sampling $p$ and $\ell$ is negligible (\crefrange{alg:heavy-hitters:line:p}{alg:heavy-hitters:line:ell}). Constructing the functions $g, h$ (\cref{alg:heavy-hitters:line:gh}) takes time $O(|A| + |B| + p)$, and computing their convolution~$f$ (\cref{alg:heavy-hitters:line:f}) takes time $O(p \log p)$ by FFT. Finally, running \cref{lem:good-approx} and assigning the estimates (\crefrange{alg:heavy-hitters:line:good}{alg:heavy-hitters:line:assign}) takes time $O((|H| + |A| \, |B| / \sigma) \cdot \log N)$. Thus, the time per round is $O(|A| + |B| + (|H| + |A| \, |B| / \sigma) \cdot \log N)$. In total, over all $O(\log N)$ rounds, this is as claimed.

\subsubsection{Correctness Analysis} \label{sec:algo:sec:heavy:sec:correctness}
Let us finally analyze the correctness. We assume now that the randomness over $p$ has already been resolved; condition on the high-probability event that the algorithm from \cref{lem:good-approx} is correct. Fix any $z \in G$ and write $k := z \bmod p$. We show that, over the remaining randomness of $\ell$, Properties~(1) and~(2) from \cref{lem:heavy-hitters} hold. Note that:
\begin{align*}
    f(k)
    &= \sum_{\substack{i, j \in \Int/p\Int\\i+j\equiv k \mod p}} g(i) \cdot h(j) \\
    &= \sum_{\substack{i, j \in \Int/p\Int\\i+j\equiv k \mod p}} \parens*{\sum_{\substack{x \in A\\x \equiv i \mod p}} \omega^{\ell x}} \cdot \parens*{\sum_{\substack{y \in B\\y \equiv j \mod p}} \omega^{\ell y}} \\
    &= \sum_{\substack{i, j \in \Int/p\Int\\i+j\equiv k \mod p}} \parens*{\sum_{\substack{x \in A\\x \equiv i \mod p}} \sum_{\substack{y \in B\\y \equiv j \mod p}} \omega^{\ell (x + y)}} \\
    &= \sum_{\substack{z' \in [0, 2N]\\z' \equiv k \mod p}} r_{A+B}(z') \cdot \omega^{\ell z'}.
\end{align*}
Therefore,
\begin{align*}
    \Ex[r(z)]
    &= \Ex[\omega^{-\ell z} \cdot f(k)] \\
    &= \Ex\brackets*{\omega^{-\ell z} \cdot \sum_{\substack{z' \in [0, 2N]\\z' \equiv k \mod p}} r_{A+B}(z') \cdot \omega^{\ell z'}} \\
    &= r_{A+B}(z) + \sum_{\substack{z \neq z' \in [0, 2N]\\z' \equiv k \mod p}} r_{A+B}(z') \cdot \Ex\brackets*{\omega^{\ell (z'-z)}} \\
    &= r_{A+B}(z),
\end{align*}
using \cref{fact:random-root-of-unity}. For this fact to apply note that $|z - z'| < 2N + 1 = u$ and therefore $z - z' \equiv 0 \mod u$ if and only if $z = z'$. Next we focus on the variance. Consider the following calculation, where complex conjugation is denoted by overlines:
\begin{align*}
    \Var[r(z)]
    &= \Ex[\, |r(z)|^2 \,] - |\Ex[\, r(z) \,]\,|^2 \\
    &= \Ex[\, r(z) \, \overline{r(z)} \,] - r_{A+B}(z)^2 \\
    &= \Ex[\, f(k) \, \overline{f(k)} \,] - r_{A+B}(z)^2 \\
    &= \sum_{\substack{z_1, z_2 \in \Int\\z_1 \equiv z_2 \equiv k \mod p}} r_{A+B}(z_1) \cdot r_{A+B}(z_2) \cdot \Ex\brackets*{\omega^{\ell (z_1 - z_2)}} - r_{A+B}(z)^2 \\
    &= \sum_{\substack{z_1 \in \Int\\z_1 \equiv k \mod p}} r_{A+B}(z_1)^2 - r_{A+B}(z)^2 \\
    &= \sum_{\substack{z \neq z' \in \Int\\z' \equiv k \mod p}} r_{A+B}(z')^2.
\end{align*}
Now we use that $z \in G$ is $\sigma$-good twice. First, each individual term $r_{A+B}(z')$ can be upper bounded by $\sigma$. By a second application, the variance is indeed bounded by
\begin{equation*}
    \Var[r(z)] = \sum_{\substack{z \neq z' \in \Int\\z' \equiv k \mod p}} r_{A+B}(z')^2 \leq \sigma \cdot \sum_{\substack{z \neq z' \in \Int\\z' \equiv k \mod p}} r_{A+B}(z') \leq \sigma^2.
\end{equation*}

Recall that we have conditioned on the event $\mathcal E$ that \cref{lem:good-approx} succeeds. Let us finally remove that condition. Recall that $\mathcal E$ happens with probability $1 - (N / \rho)^{-10}$. If $\mathcal E$ fails, then $r(z)$ is not necessarily an unbiased estimator. However, even then $r(z)$ differs from $r_{A + B}(z)$ by at most $N$, and thus the total effect on the expectation is at most $\Pr(\overline{\mathcal E}) \cdot N \leq \rho$. Similarly, if $\mathcal E$ fails then the variance is at most $N^2$, and thus the effect on the variance is at most
\begin{equation*}
    \Var[r(z)] \leq \Var[r(z) \mid \mathcal E] + \Pr(\overline{\mathcal E}) \cdot O(N^2) \leq \Var[r(z) \mid \mathcal E] + O(1/N).
\end{equation*}
The first term is at most $\sigma^2$ as before, and the second term can also be absorbed by rescaling $\sigma$.

\subsubsection{Computing with Complex Numbers} \label{sec:algo:sec:heavy:sec:complex}
It remains to replace the complex-valued outputs by real-valued outputs, and to remove the assumption that all computations have infinite precision. For the first issue, we can simply discard the imaginary parts for all outputs. Since the expected values $\Ex[r(z)]$ are real-valued, this does not affect Property~(1), and since the variance can only decrease, this also does not violate Property~(2).

For the second issue, we represent complex numbers $a + bi \in \Complex$ as pairs $(a, b)$, where both components are stored with fixed-point precision in $L = \Theta(\log(N / \rho))$ bits. That is, we represent complex numbers with magnitude up to $N^{O(1)}$ with absolute error at most $\rho^{O(1)}$. This precision error affects the resulting estimates with an absolute error of $\pm \rho$. This affects the expectation as claimed in the lemma statement; the contribution to the variance can be absorbed by rescaling $\sigma$ (using that $\rho$ is smaller and $\sigma$ is larger than any absolute constants without loss of generality).

This implementation affects the running time as follows. Each arithmetic operation can be implemented in time $L^{O(1)}$. Moreover, it is known that the convolution of two complex-valued functions can be computed without a substantial blow-up in the required precision; specifically, computing the convolution of two length-$N$ sequences with $L$ bits of precision takes time $O(N \log N L^{O(1)})$; see, e.g.,~\cite{Schonhage82}. (In fact, we will ultimately set $L = O(\log n)$, and for this choice the overhead~$L^{O(1)}$ can be avoided in the word RAM model altogether.) Finally, we need to prepare the root of unity $\omega$ in this representation; this can be done in time $L^{O(1)}$ e.g., by evaluating the Taylor expansions of the cosine and sine functions to obtain the real and imaginary parts, respectively. This completes the proof of \cref{lem:heavy-hitters}.

\subsection{Estimating Sumsets}
We will now complete the proof of \cref{thm:sumset-estimate}, i.e., we will extend the previously established estimator for the heavy hitters to a full estimator for sumset multiplicities $r_{A+B}$.

To estimate the missing light entries, it suffices to use naive sampling. However, we need to do so in two stages: first to classify which elements are heavy hitters and which are light, and second to obtain the light estimates. Both proofs are very similar, though require somewhat incomparable guarantees.

\begin{lemma}[Heavy/Light Classification] \label{lem:heavy-light}
There is a randomized algorithm that, given sets $A, B \subseteq [0, N]$ and parameters $\sigma, \delta > 0$, computes a partition $[0, 2N] = L \sqcup H$ into \emph{light} and \emph{heavy} elements such that $|H| \leq O(|A| \, |B| / \sigma)$ and:
\begin{enumerate}[label=(\roman*)]
    \item for all $z \in L$, $r_{A+B}(z) \leq \sigma$, and
    \item for all $z \in H$, $r_{A+B}(z) \geq \frac{1}{2} \cdot \sigma$.
\end{enumerate}
The algorithm is correct with probability $1 - \delta$ and runs in time $O(|A| + |B| + |A| \, |B| / \sigma \cdot \log(N/\delta))$.
\end{lemma}
\begin{proof}
Let $R = \Theta(\log (N / \delta))$ and $q = R \cdot |A| \, |B| / \sigma$, and draw uniform samples $a_1, \dots, a_q \in A, b_1, \dots, b_q \in B$. Write $z_i = a_i + b_i$. Let $H$ be the set of elements $z$ that appear at least $\frac{3}{4} \cdot R$ times among the samples $z_1, \dots, z_q$. The expected number of samples equal to $z$ is
\begin{equation*}
    q \cdot \frac{r_{A+B}(z)}{|A| \, |B|} = R \cdot \frac{r_{A+B}(z)}{\sigma}.
\end{equation*}
Thus, if $r_{A+B}(z) \geq \sigma$ the expected number of samples is at least $R$, and by Chernoff's bound the number of samples is at least $\frac{3}{4} \cdot R$ with high probability $1 - \delta / (2N+1)$. Similarly, if $r_{A+B}(z) \leq \frac{1}{2} \cdot \sigma$, then the expected number of samples is at most $\frac{1}{2} \cdot R$, and by Chernoff's bound the number of samples is less than $\frac{3}{4} \cdot R$ with high probability. Here, in both cases ``with high probability'' means with probability $1 - \delta / (2N + 1)$ when choosing the hidden constant in $R$ sufficiently large; then a union bound shows that the overall success probability is still at least $1 - \delta$.
\end{proof}

We remark that a derandomization of the previous lemma is known~\cite{Fischer0X25}, at the cost of an $N^{o(1)}$-loss in the running time.

\begin{lemma}[Light Estimator] \label{lem:light-estimate}
There is a randomized algorithm that, given sets $A, B \subseteq [0, N]$ and a parameter $\sigma > 0$, computes estimates $r$ such that:
\begin{enumerate}[label=(\arabic*)]
    \item $\Ex[r(z)] = r_{A+B}(z)$ for all $z \in [0, 2N]$, and
    \item $\Var[r(z)] \leq r_{A+B}(z) \cdot \sigma$ for all $z \in [0, 2N]$, and
    \item at most $O(|A| \, |B| / \sigma)$ estimates $r(z)$ are nonzero.
\end{enumerate}
The algorithm runs in expected time $O(|A| + |B| + |A| \, |B| / \sigma)$.
\end{lemma}
\begin{proof}
Again, draw $q$ independent uniform samples $a_1, \dots, a_q \in A$ and $b_1, \dots, b_q \in B$, and let $z_i = a_i + b_i$. Let $r_i$ denote the function mapping $z_i$ to $|A| \, |B| / q$, and all other inputs to $0$. Compute $r = \sum_i r_i$. Then~$r$ is an unbiased estimator of $r_{A+B}$ because, for any $z$,
\begin{equation*}
    \Ex[r(z)] = \sum_{i=1}^q \Ex[r_i(z)] = \sum_{i=1}^q \frac{|A| \, |B|}{q} \cdot \frac{r_{A+B}(z)}{|A| \, |B|} = r_{A+B}(z).
\end{equation*}
Similarly,
\begin{equation*}
    \Var[r(z)] \leq \sum_{i=1}^q \Var[r_i(z)] \leq \sum_{i=1}^q \Ex[r_i(z)^2] = \sum_{i=1}^q \frac{|A|^2 |B|^2}{q^2} \cdot \frac{r_{A+B}(z)}{|A| \, |B|} = r_{A+B}(z) \cdot \sigma.
\end{equation*}
Finally, consider the running time. After preprocessing $A$ and $B$ in linear time $O(|A| + |B|)$, we can sample the $a_i$'s and $b_i$'s in constant time each. Thus, the computation of $r$ takes time $O(q)$.
\end{proof}

\begin{proof}[Proof of \cref{thm:sumset-estimate}]
We start with the algorithm. We may assume $\sigma \geq 1$ (otherwise enumerate $A \times B$) and replace $\rho$ by $\min\{\rho, 1\}$. We run the heavy/light decomposition (\cref{lem:heavy-light}) with some parameter $\delta$. Let $\mathcal E$ denote the event that the algorithm succeeds; the probability of $\mathcal E$ is at least $1 - \delta$ and we condition on this event for the moment. Denote the resulting sets by $H$ and $L$.

We will now separately assign $r(z)$ for the values in $H$ and $L$. The values for $z \in H$ are assigned by one call to \cref{lem:heavy-hitters}. For the light ones, we instead run \cref{lem:light-estimate} with parameter $\sigma$ to compute some estimate~$r'$ and choose $r(z) := r'(z)$ for all $z \in L$. For both cases we have $\Ex[r(z) \mid H] = r_{A+B}(z) \pm \rho$. In the heavy case the variance bound $\Var[r(z)] \leq \sigma^2$ is clear, and in the light case it follows from $\Var[r(z)] \leq r_{A+B}(z) \cdot \sigma$ since $r_{A+B}(z) \leq \sigma$ for all $z \in L$. This proves Properties~(1) and~(2). For Property~(3), we observe that the support of $r$ is upper bounded by $|H|$ plus the support of the light estimator; both are at most $O(|A| \, |B| / \sigma)$ by \cref{lem:heavy-light,lem:light-estimate}.

Let us now remove the condition on the event $\mathcal E$. Irrespective of $\mathcal E$, $r(z)$ is an unbiased estimator (up to the precision error $\rho$), so Property~(1) is not affected. Only the variance bound in Property~(2) may be affected. For any fixed choice of $H$ and $L$, the conditional variance is at most $\sigma^2 + O(N^2)$, even if $\mathcal E$ fails. Since all conditional means lie within $\rho$ of $r_{A+B}(z)$, the law of total variance gives
\begin{equation*}
    \Var[r(z)] \leq \Ex[\Var[r(z) \mid H]] + \rho^2 \leq \sigma^2 + O(\delta N^2) + \rho^2.
\end{equation*}
Taking $\delta = 1/N^2$, the extra terms are $O(1)$ and can be absorbed into $\sigma^2$ by rescaling $\sigma$, since $\rho \leq 1 \leq \sigma$.

Finally, we analyze the running time. It is dominated by the calls to \cref{lem:heavy-light,lem:light-estimate,lem:heavy-hitters}, and the running times of all three lemmas can be bounded by $(|A| + |B| + |H| + |A| \, |B| / \sigma) \cdot \log^{O(1)} (N/\rho)$. \cref{lem:heavy-light} gives that $|H| \leq O(|A| \, |B| / \sigma)$ (unconditionally), therefore, the contribution of $|H|$ to the running time is dominated.
\end{proof}

\subsection{Estimating Sumsets of Aligned Intervals}
For our application to Text-to-Pattern Hamming Distances, it is not sufficient to estimate sumsets of \emph{sparse} sets $A, B$. In general, we support estimating sumsets of sets $A$ and $B$ that can be expressed as the union of few intervals (i.e., with small run-length encoding of their indicator strings). Let us call an interval \emph{aligned} if its starting position is a multiple of its length. The following lemma is sufficient:

\begin{lemma} \label{lem:conv-interval-estimate}
There is a randomized algorithm that, given sets $A, B \subseteq [0, N]$ that are represented as the union of $k_A$ and $k_B$ aligned intervals of lengths $L_A$ and $L_B$, respectively, $L_A \mid L_B$, and parameters $\sigma, \rho > 0$, computes estimates~$r$ such that
\begin{enumerate}[label=(\arabic*)]
    \item $\Ex[r(z)] = r_{A+B}(z) \pm \rho$ for all $z \in [0, 2N]$,
    \item $\Var[r(z)] \leq \sigma^2$ for all $z \in [0, 2N]$, and
    \item $r$ is a piece-wise linear function represented by an expected number of $O(k_A k_B \sqrt{L_A L_B} / \sigma)$ linear pieces.
\end{enumerate}
The algorithm runs in expected time $(k_A + k_B L_B / L_A + k_A k_B \sqrt{L_A L_B} / \sigma) \cdot (\log (N/\rho))^{O(1)}$.
\end{lemma}
\begin{proof}
Let $B' \subseteq \Int$ denote the starting positions of the aligned intervals in $B$ in the sense that $y \in B'$ if and only if a $B$-interval starts at position $y L_B$. Similarly, for each $0 \leq i < L_B / L_A$, let $A'_i$ denote the set of starting positions in $A$ that are equal to $i L_A$ modulo $L_B$; that is, $x \in A'_i$ if and only if there is an $A$-interval starting at position $x L_B + i L_A$. In other words, we have expressed
\begin{equation*}
    A = \bigsqcup_{0 \leq i < L_B/L_A} (A'_i \cdot L_B + i \cdot L_A + [L_A]) \quad\text{and}\quad B = B' \cdot L_B + [L_B].
\end{equation*}
We apply \cref{thm:sumset-estimate} $L_B / L_A$ times, each with the same parameters $\sigma'$ and $\rho$, to estimate $r_{A_i' + B'}$ by $r_i$. For convenience, assume that out-of-bounds values of $r_i$ are zero. From these estimates we interpolate an estimate $r$ of $r_{A+B}$ as follows. For each $z$, written as $z = z_1 L_B + z_2$ for $0 \leq z_2 < L_B$, let
\begin{equation*}
    r(z) := \sum_{0 \leq i < L_B / L_A} \sum_{c=0}^1 r_i(z_1 - c) \cdot r_{[L_A] + [L_B]}(z_2 + cL_B - i L_A).
\end{equation*}

We claim that this choice indeed satisfies the three properties in the lemma statement. To see this, express $r_{A+B}(z)$ as 
\begin{equation*}
    r_{A+B}(z) = \sum_{0 \leq i < L_B / L_A} \abs{\set{(x_1, x_2, y_1, y_2) \in A_i' \times (i L_A + [L_A]) \times B' \times [L_B] : x_1 L_B + x_2 + y_1 L_B + y_2 = z}}.
\end{equation*}
This condition is equivalent to the two equations $z_1 = x_1 + y_1 + c$ and $z_2 = x_2 + y_2 - c L_B$ for some carry bit $c \in \set{0, 1}$. Hence, $x_1 + y_1 = z_1 - c$ and $(x_2 - iL_A) + y_2 = z_2 + cL_B - iL_A$, where $x_2 - iL_A \in [L_A]$. Thus,
\begin{equation*}
    r_{A+B}(z) = \sum_{0 \leq i < L_B / L_A} \sum_{c=0}^1 r_{A'_i + B'}(z_1 - c) \cdot r_{[L_A] + [L_B]}(z_2 + cL_B - i L_A).
\end{equation*}
It follows that $\Ex[r(z)] = r_{A+B}(z) \pm N \rho$, proving~(1) by rescaling $\rho$. To bound the variance, note that $r(z)$ is the sum of at most~$L_B / L_A$ pairs of terms $r_i(z_1 - 1)$ and $r_i(z_1)$, multiplied by some integer at most $L_A$. Each such term has variance at most~$L_A^2 \cdot (\sigma')^2$. The covariances across different $i$'s are zero (since these are obtained independently), and, based on the fact that $\Cov[X, Y] \leq 2 \Var[X] + 2\Var[Y]$, the covariance between any pair $r_i(z_1 - 1)$ and~$r_i(z_1)$, possibly multiplied with integers up to $L_A$, is at most $O(L_A^2 \cdot (\sigma')^2)$. Therefore, the total variance is $\Var[r(z)] \leq O(L_B / L_A \cdot L_A^2 \cdot (\sigma')^2)$. Choosing $\sigma' = \Theta(\sigma / \sqrt{L_A L_B})$ for some sufficiently small hidden constant, the variance becomes $\Var[r(z)] \leq \sigma^2$, proving~(2).

It remains to prove~(3), i.e., to bound the number of linear segments required to represent $r$. To this end, note that $r_{[L_A] + [L_B]}$ can be represented by four linear segments: this function has a trapezoid shape, with a first slope-$1$ segment from $0$ to $L_A-1$, a second constant segment, and third slope-$-1$ segment from $L_B - 1$ to $L_A + L_B - 1$, and a final constant segment (with value $0$). The function $r$ can be expressed as the sum of several shifted copies of $r_{[L_A] + [L_B]}$, where each nonzero entry in $r_i$ leads to at most two copies. Thus, by \cref{thm:sumset-estimate} the expected number of linear segments is bounded by
\begin{equation*}
    O\parens*{\sum_{0 \leq i < L_B / L_A} \Ex[\nnz(r_i)]} \leq O\parens*{\sum_{0 \leq i < L_B / L_A} \frac{|A_i'| \, |B'|}{\sigma'}} = O\parens*{\frac{k_A k_B \sqrt{L_A L_B}}{\sigma}}.
\end{equation*}

Finally, consider the running time. The sets $A'_i$ and $B'$ can be constructed in time $O(k_A + k_B)$. Computing the estimates $r_i$ takes time $(|A'_i| + |B'| + |A'_i| \, |B'| / \sigma') \cdot (\log(N/\rho))^{O(1)}$ by \cref{thm:sumset-estimate}. In total, these calls take time $(k_A + k_B L_B / L_A + k_A k_B \sqrt{L_A L_B} / \sigma) \cdot (\log(N/\rho))^{O(1)}$, using that $\sum_i |A_i'| \leq k_A$ and that $\sum_i |B'| \leq k_B L_B / L_A$. Finally, the function $r$ can be assembled in piece-wise linear representation in time proportional to the number of linear pieces, up to logarithmic factors; the expected number of pieces, as argued before, is $O(k_A k_B \sqrt{L_A L_B} / \sigma)$.
\end{proof}
\section{Approximate Text-to-Pattern Hamming Distances}
In this section we finally prove \cref{thm:main}, by giving our $\tilde O(n \epsilon^{-2/3})$-time algorithm for $(1+\epsilon)$-approximate Text-to-Pattern Hamming Distances.

The first step is to apply the following reduction (which we reprove in Appendix \ref{sec:proveRLEappendix} for convenience):

\begin{restatable}[{\cite{GawrychowskiU18,Chan0WX23}}]{theorem}{reducetoRLE}
\label{thm:reducetoRLE}
Solving the $(1 + \epsilon)$-approximate Text-to-Pattern Hamming Distance problem reduces in time $\tilde O(n)$ to $O(\log n)$ instances
of solving \emph{additive} $\epsilon k$-approximate Text-to-Pattern Hamming Distance on two strings of length $O(n)$ and run-length encoding at most $O(k)$.

\end{restatable}

Focus on one instance constructed by the theorem with text $T$ and pattern $P$. We assume that $|T|$ and $|P|$ are powers of $2$ (with padding if necessary) and that $|T| = \Theta(|P|)$. Instead of approximating the Hamming distances, we approximate the number of \emph{matches}
\begin{equation*}
    M_{T, P}(i) := \abs{\set{j \in [|P|] : T[i + j] = P[j]}}
\end{equation*}
for each shift $i \in [0, |T| - |P|]$ that aligns the pattern with the text. The Hamming distance at shift $i$ can then be recovered as $|P|$ minus the number of matches, without worsening the additive approximation error.

We now compute a partition of $[|T|]$ based on the text $T$. Recall that a \emph{dyadic interval} is an interval of the form $[j 2^\ell, (j + 1)2^\ell)$, where $j$ and $\ell$ are integers. A basic fact is that every interval in the range $[N]$ can be partitioned into at most $2 \ceil{\log N}$ consecutive dyadic intervals. We now consider each run in $T$, and partition the corresponding interval into at most $2 \log n$ dyadic intervals. Overall this results in a partition of~$[|T|]$ into at most $O(k \log n)$ dyadic intervals. We similarly partition $[|P|]$ via the pattern $P$. Now enumerate all pairs $L_A, L_B$ of powers of $2$ in the ranges $[|T|]$ and $[|P|]$. Let $T'$ be the string obtained from $T$ that retains the characters in the dyadic intervals of length $L_A$, and let $P'$ be the string obtained from $P$ that retains the characters in the dyadic intervals of length $L_B$; replace the deleted characters by some dummy symbol that cannot participate in any matches. To approximate the number of matches between $P$ and $T$, it suffices to approximate the number of matches between $P'$ and $T'$ over all $O(\log^2 n)$ instances. The total additive error worsens by $O(\log^2 n)$, which can be absorbed by rescaling $\epsilon$.

For the remainder of this section focus only on $T'$ and $P'$. Assume that $L_A \leq L_B$; the other case is symmetric. We enumerate all characters $s$ that appear in~$T'$ or~$P'$. Let
\begin{align*}
    A_s &= \set{i : T'[i] = s}, \\
    B_s &= \set{j : P'[|P| - 1 - j] = s}
\end{align*}
indicate the positions of that character in the text and in the \emph{reversed} pattern, respectively. By the previous paragraph $A_s$ is the union of aligned intervals of length $L_A$; let $k_{A, s}$ denote their number. Similarly, $B_s$ is the union of $k_{B, s}$ aligned intervals of length $L_B$. In this notation, the total number of matches at shift $i$ can be expressed exactly as
\begin{equation*}
    M_{T', P'}(i) = \sum_s \abs{\set{j \in [|P|] : T'[i + j] = P'[j] = s}} = \sum_s r_{A_s + B_s}(|P| - 1 + i).
\end{equation*}
In particular, to estimate the number of matches, it suffices to estimate the sumset multiplicities $r_{A_s + B_s}$ for all symbols $s$. We do so in two cases:
\begin{itemize}
    \item \emph{(Heavy)} If $k_{B, s} \geq t$ for some threshold $t$ to be chosen, then we compute $r_{A_s + B_s}$ by one FFT. Naively this takes time $\tilde O(n)$. However, we can improve this using that both sets can be seen as unions of intervals of length $L_A$. We therefore only need to compute the convolution of the sets consisting of the starting points and can interpolate between the aligned points. The running time is $\tilde O(n / L_A)$.
    \item \emph{(Light)} Otherwise we estimate $r_{A_s + B_s}$ by \cref{lem:conv-interval-estimate} with parameters \smash{$\sigma_s := \epsilon \sqrt{k k_{B, s}}$} and $\rho = n^{-10}$.
\end{itemize}
Let $r_s$ denote our estimate of $r_{A_s + B_s}$ and let $r = \sum_s r_s$ denote the final estimates. We show that $r(z)$ is an additive $\epsilon k$-approximation of $\sum_s r_{A_s + B_s}(z)$, with probability at least $2/3$. First, both cases give unbiased estimators (up to the precision error $\rho$), and thus
\begin{equation*}
    \Ex[r(z)] = \sum_s \Ex[r_s(z)] = \sum_s (r_{A_s + B_s}(z) \pm \rho) \leq \sum_s r_{A_s + B_s}(z) + O(n \rho).
\end{equation*}
We pick $\rho = n^{-10}$ so that the error term becomes negligible. Now consider the variance. The estimates $r_s$ obtained in the heavy case are deterministic and thus have zero variance. The estimates from the light case contribute:
\begin{equation*}
    \Var[r(z)] \leq \sum_{\text{$s$ light}} \sigma_s^2 \leq \epsilon^2 k \sum_{\text{$s$ light}} k_{B, s} \leq \epsilon^2 k^2.
\end{equation*}
Thus, Chebyshev's inequality implies that, with probability at least $2/3$,
\begin{equation*}
    \abs*{r(z) - \sum_s r_{A_s + B_s}(z)} \leq O(\epsilon k).
\end{equation*}
By rescaling $\epsilon$ by a constant this becomes $\epsilon k$ as claimed. And by repeating this algorithm independently $\Theta(\log n)$ times and selecting the median estimate per $z$, the error probability becomes $n^{-100}$. This completes the description of the algorithm and the correctness analysis.

It remains to analyze the running time. We can enumerate all symbols $s$ in time $O(n)$. In total, there can be at most $O(n / (t L_B))$ symbols that fall into the heavy case (each such symbol takes up at least $t L_B$ characters in the pattern). Thus, the total time of the heavy case is
\begin{equation*}
    O\parens*{\frac{n^2}{t L_A L_B}}.
\end{equation*}
By \cref{lem:conv-interval-estimate} the running time of the light case can be bounded as follows per $s$, ignoring $\log n$ factors for convenience:
\begin{equation*}
    O\parens*{k_{A, s} + \frac{k_{B, s} L_B}{L_A} + \frac{k_{A, s} k_{B, s} \sqrt{L_A L_B}}{\sigma_s}}.
\end{equation*}
Over all symbols $s$, we can bound $\sum_s k_{A, s} \leq k \leq n$ and $\sum_s k_{B, s} \leq O(n / L_B)$, so the total time is
\begin{align*}
    &O\parens*{n + \sum_{\text{$s$ light}} \frac{k_{A, s} k_{B, s} \sqrt{L_A L_B}}{\sigma_s}} \\
    &\qquad= O\parens*{n + \sum_{\text{$s$ light}} \frac{k_{A, s} \sqrt{k_{B, s} L_A L_B}}{\epsilon \sqrt{k}}} \\
    &\qquad= O\parens*{n + \frac{k\sqrt{t L_A L_B}}{\epsilon \sqrt{k}}} \\
    &\qquad= O\parens*{n + \frac{\sqrt{n t L_A L_B}}{\epsilon}}.
\end{align*}
We choose
\begin{equation*}
    t := \frac{\epsilon^{2/3} n}{L_A L_B}
\end{equation*}
to balance both contributions; the total time becomes $\tilde O(n \epsilon^{-2/3})$. Finally, we also need to compute the full function $r$ by summing over all $r_s$, but naively spending time $O(n)$ per light iteration would be too slow. To this end, we use that each estimated function $r_s$ is represented by an expected number of $O(k_{A, s} k_{B, s} \sqrt{L_A L_B} / \sigma_s)$ linear pieces. Using an appropriate range tree data structure, we can thus perform all updates of the form $r \gets r + r_s$ in time proportional to the number of linear pieces, up to log-factors. Hence, this running time can be charged to the running time of the light case as analyzed before. This completes the proof of \cref{thm:main}.
\section{Matching Lower Bound for a Restricted Class of Algorithms} \label{sec:hard-instance}
In this section we show that the running time $\tilde O(\epsilon^{-2/3} n)$ is best-possible for a certain natural class of algorithms which we call \emph{symbolwise} algorithms. Before we make this class precise, let us first describe the hard instance. Throughout, focus on the simpler problem to compute an \emph{additive} $\epsilon n$-approximation of all Hamming distances, and focus on the parameter regime $\epsilon = n^{-3/4}$. That is, our goal is to compute an additive $n^{1/4}$-approximation of all Hamming distances, or equivalently, of the number of matches per alignment.

Consider the following instance. Let $\Sigma$ be an alphabet of size $n^{1/2}$. We also include some dummy symbol~$\bot$ into $\Sigma$ that will never participate in matches, and when we \emph{delete} a character in $T$ we mean that it is replaced by $\bot$. Let $n = 2m$. Let $T \in \Sigma^n$ be a uniformly random string where we delete each symbol with probability $1/2$, and let $P \in \Sigma^m$ be a uniformly random string. In this instance, each symbol is expected to contribute about $1$ match to each shift, and thus the total number of matches per shift is expected to be~$\sqrt n$. The variance is also $\sqrt n$, and therefore the standard deviation is $n^{1/4}$. In this regime we are forced to consider the contribution of every character. But as each character contributes, say, $0$ or $1$, we are either forced to spend time $\Omega(1)$ per character and per shift, or we risk estimating the contribution which will lead to prohibitively large estimation errors. In the former case we spend time $\Omega(n^{3/2})$.

Defining the class of algorithms is technically tedious. Informally, we say that an algorithm is symbolwise if it considers the input symbols $s \in \Sigma$ \emph{separately}, computes estimates $r_s$ of the number of matches as a sparsely represented function, and aggregates these estimates~$r_s$. Formally:

\begin{definition}[Symbolwise Algorithms]
Let $\mathcal A$ be an algorithm that receives as input sets $A, B$ and reports a sparsely represented function $r : \Int \to \Real$. Then consider the following algorithm $\mathcal S$ for the approximate Text-to-Pattern Hamming Distances problem with input text $T$ and input pattern $P$:
\begin{enumerate}
    \item Initialize $r : [0, n - m] \to \Real$ as the all-zero function.
    \item Enumerate all symbols $s \in \Sigma$ that appear in text and pattern. For each one, let $A_s = \set{i : T[i] = s}$ and $B_s = \set{j : P[j] = s}$, and call algorithm $\mathcal A$ on the sets $A_s, B_s$ to compute some sparsely represented function $r_s$. Then, for each nonzero entry $r_s(z)$, update $r(z) := r(z) + r_s(z)$.
    \item Return $r$ as the estimated number of matches (i.e., $r(z)$ approximates $m - \HD(T\intervalcc{z}{z+m-1} , P)$).
\end{enumerate}
We call $\mathcal S$ a \emph{symbolwise} algorithm.
\end{definition}

That is, we force the algorithm $\mathcal S$ to consider the symbols separately by making $\mathcal A$ see only one symbol at a time. This class contains the classical exact algorithms by Fischer and Paterson~\cite{FischerP74} and by Abrahamson~\cite{Abrahamson87}. Applied to the specific hard instance mentioned above, in the critical regime~\makebox{$\epsilon = n^{-3/4}$}, the class also contains our algorithm, as well as the predecessor by Chan, Jin, Vassilevska~W., and Xu~\cite{Chan0WX23}. Indeed, in both cases the algorithms essentially degenerate to computing one FFT or solving one Sumset Estimation per symbol, without any detours through run-length encoded strings (after all, random strings do not admit run-length compressions). However, the class does not contain the projection-based~\cite{Karloff93,KopelowitzP15,KopelowitzP18} and sampling-based~\cite{Indyk98a,ChanGKKP20} algorithms.

We prove that any symbolwise algorithm cannot beat the $\epsilon^{-2/3}$ barrier.

\begin{lemma} \label{lem:symbolwise}
Any symbolwise algorithm that computes an \emph{additive} $\epsilon n$-approximation that is correct with probability at least $2/3$ per alignment requires time $\Omega(\epsilon^{-2/3} n)$ in the regime $\epsilon = \Theta(n^{-3/4})$.
\end{lemma}

The following simple consequence of the Kolmogorov--Rogozin inequality~\cite{Rogozin61} will be helpful:

\begin{lemma}[Kolmogorov--Rogozin] \label{lem:kolmogorov-rogozin}
Let $X_1, \dots, X_n \in \set{0, 1}$ be independent random variables, and let $X = \sum_{i=1}^n X_i$. Then, for any interval $I$ of length $L \geq 1$,
\begin{equation*}
    \Pr(X \in I) \leq C L \parens*{\sum_{i=1}^n \min\set{\Pr(X_i = 0), \Pr(X_i = 1)}}^{-1/2}.
\end{equation*}
Here, $C$ is an absolute constant.
\end{lemma}

\begin{proof}[Proof of \cref{lem:symbolwise}]
Let $\mathcal S$ be a symbolwise algorithm, and let $\mathcal A$ be the corresponding algorithm operating on a single symbol. Let $\Sigma$ be an alphabet of size $\Theta(n^{1/2})$, let $\bot \not\in \Sigma$ denote some dummy symbol, and write $\overline\Sigma = \Sigma \cup \set{\bot}$. Let~$T \in \Sigma^n$ and $P \in \Sigma^m$, for $n = 2m$, be uniformly random strings. Let $\overline T \in \overline\Sigma^n$ be the string obtained from~$T$ by deleting each symbol independently with probability $1/2$. We consider the execution of $\mathcal S$ on text~$\overline T$ and pattern $P$, with some parameter $\epsilon = c n^{-3/4}$ for some sufficiently small constant~$c$.

A lower bound on the running time of $\mathcal S$ is
\begin{equation*}
    \Omega\parens*{\sum_s \nnz(r_s)},
\end{equation*}
since each nonzero entry of a function $r_s$ requires an update that takes time $\Omega(1)$. We will argue that the total number of nonzeroes across $r_s$ must be $\Omega(n^{3/2})$.

Fix any shift $z \in [0, n - m]$; it suffices to show that the number of nonzero entries $r_s(z)$ is at least $\Omega(n^{1/2})$. First, focus on the randomness of $T$ and $P$. Let $S \subseteq \Sigma$ be the set of characters with exactly one match at position $z$, i.e., let
\begin{equation*}
    M_s = \abs{\set{j : T[z + j] = P[j] = s}},
\end{equation*}
and
\begin{equation*}
    S = \set{s \in \Sigma : M_s = 1}.
\end{equation*}
Each $s \in \Sigma$ is placed in $S$ with probability $\Omega(1)$: the conditions $T[z + j] = P[j] = s$ can be seen as independent Bernoulli variables with probability $1/|\Sigma|^2$, and thus $M_s$ behaves like a binomial random variable  sampled from $\operatorname{Bin}(m, 1/|\Sigma|^2)$. For our choice of parameters, the probability that $M_s = 1$ is $\Omega(1)$. Thus, $\Ex[|S|] = \Omega(|\Sigma|)$, and Markov's inequality (applied on $|\Sigma| - |S|$) implies that with probability $\Omega(1)$ we have $|S| \geq \Omega(|\Sigma|)$. In this event we call $z$ \emph{relevant}, and we will only focus on the relevant shifts $z$ in the following.

Now fix the randomness of $T$ and $P$, and only leave the randomness of $\overline T$, i.e., of deleting symbols, unresolved. To match our notation, let
\begin{align*}
    A_s &= \set{i : T[i] = s}, \\
    \overline A_s &= \set{i : \overline T[i] = s}, \\
    B_s &= \set{i : P[i] = s}.
\end{align*}
Moreover, write
\begin{equation*}
    \overline M_s = \abs{\set{j : \overline T[z + j] = P[j] = s}},
\end{equation*}
The sets $A_s$ and $B_s$ are fixed, and $\overline A_s \subseteq A_s$ is a uniformly random subset. In particular, each character $s \in S$ contributes either $0$ or $1$ match between $\overline T[z .. z + m - 1]$ and $P$, and thus the random variables $\overline M_s$ are independent Bernoulli variables for all $s \in S$.

Now consider the functions $r_s$ returned by the algorithm for the symbols $s \in S$. View $r_s(z)$ as a random variable over the randomness of $\overline T$ and over the randomness of the algorithm $\mathcal A$. Then $r_s(z)$ may be dependent on $\overline M_s$ (in fact, in any reasonable algorithm it \emph{should} be dependent), but $r_s(z)$ is independent from $\overline M_t$ for any other symbol $t \neq s$ (since $\mathcal A$ only receives as input $\overline A_s$ and $B_s$, but not $\overline A_t$). Let $D_s = \overline M_s - r_s(z)$, and let $D = \sum_{s \in S} D_s$.

Let us call $s$ \emph{cheap} if $\Pr(r_s(z) = 0) \geq 3/4$. If $s$ is cheap, then, intuitively, $r_s(z)$ and $\overline M_s$ cannot be as dependent on each other as they should be. Specifically, for any cheap $s$ we must have that
\begin{equation*}
    p_s := \Pr(\overline M_s = 1 \mid r_s(z) = 0) \in [1/4, 3/4].
\end{equation*}
First suppose that at least half the symbols in $S$ are cheap; denote these cheap symbols by $C \subseteq S$. Now draw the randomness of $r_s(z)$ for all symbols $s \in C$, and let $C' = \set{s \in C : r_s(z) = 0}$. By Chernoff's bound, with high probability we have $|C'| \geq |C| / 2 = \Omega(|S|) = \Omega(|\Sigma|)$. Now consider the sum
\begin{equation*}
    X = \sum_{s \in C'} D_s = \sum_{s \in C'} \overline M_s;
\end{equation*}
over the remaining randomness of $\overline M_s$, conditioning only on $r_s(z) = 0$, each term $\overline M_s$ in the sum behaves like a Bernoulli random variable with parameter $p_s \in [1/4, 3/4]$. In particular, by the Kolmogorov--Rogozin inequality (\cref{lem:kolmogorov-rogozin}), there is some interval length $L = \Omega(|C'|^{1/2}) = \Omega(n^{1/4})$ such that $X$ falls into any fixed length-$L$ interval with probability at most $1/100$. Now recall that the algorithm outputs an estimate $r(z)$ which is supposed to satisfy
\begin{equation*}
    \epsilon n \geq \abs*{\sum_{s \in \Sigma} \overline M_s - r(z)} = \abs*{\sum_{s \in \Sigma} (\overline M_s - r_s(z))} = \abs*{\sum_{s \in \Sigma} D_s},
\end{equation*}
with probability at least $2/3$. However, even when we condition on the outcome of the algorithm, the right-hand side falls into any fixed length-$L$ interval with probability at most $1/100$. We therefore must have $\epsilon n \geq L = \Omega(n^{1/4})$ or equivalently $\epsilon = \Omega(n^{-3/4})$. This contradicts our initial choice of $\epsilon$ when choosing~$c$ sufficiently small.

We must therefore be in the case that at most half of the symbols in $S$ are cheap. In other words, for at least half the symbols $s \in S$, we have $\Pr(r_s(z) \neq 0) \geq 1/4$. Chernoff's bound implies that in this case the total number of nonzero $r_s(z)$ is $\Omega(|S|) = \Omega(\sqrt{n})$ with high probability. This proves our initial claim, and completes the proof of \cref{lem:symbolwise}.
\end{proof}

\section*{AI Disclosure}
No generative AI tools were used to generate the mathematical ideas or writing of this paper. The authors used ChatGPT Astra~6 (OpenAI) and Claude Opus~5.5 (Anthropic) for proofreading and verifying correctness of the subsequent writeup. 

\bibliographystyle{alphaurl}
\bibliography{ref}

\appendix
\newcommand\runs{\text{runs}}

\section{Reducing Bounded Distance to Short Run-Length Encoding}\label{sec:proveRLEappendix}
In this section we prove \cref{thm:reducetoRLE}.

\reducetoRLE*

The main component is the following theorem, implicitly proven in \cite{GawrychowskiU18} (Theorem 10, Lemmas 6-9).

\begin{theorem}[{\cite{GawrychowskiU18}}]\label{thm:GUthm}
    Given a pattern $P$ of length $m$, text $T$ of length $n = 2m$ and integer $k\leq m$, there exists an algorithm running in time $O(m\log ^3 m)$ that computes strings $P^*,T^*$ of length $O(m)$ and RLE $O(k)$, a mapping $\phi: [m+1]\to [|T^*|-|P^*|+1]$ and a constant $c$. For each $i \in \intervalcc{0}{m}$ the algorithm does one of the following:
    \begin{itemize}
        \item Computes $\HD(T\intervalcc{i}{i + m - 1}, P)$ exactly.
        \item Discards $i$ and determines that $\HD(T\intervalcc{i}{i + m - 1}, P) > k$.
        \item Guarantees that $\HD(T\intervalcc{i}{i + m - 1}, P) = \HD(T^*\intervalcc{\phi(i)}{\phi(i) + |P^*| - 1}, P^*)-c$. 
    \end{itemize}
\end{theorem}

We first prove \cref{thm:reducetoRLE} using this result.

\begin{proof}[Proof of \cref{thm:reducetoRLE}]
    For every power of 2, $k = 1, 2, 4,\ldots, 2^{\lfloor\log m\rfloor}$, define $K = \min(2k, m)$ and run the algorithm from \cref{thm:GUthm} on $P, T, K$. Compute an additive $+\epsilon k$-approximation on the output $P^*, T^*$.

    For every index $i\in [m+1]$ let $k_i$ be the minimal value of $k$ for which the iteration of the algorithm with $T,P,K$ did not discard $i$. \cref{thm:GUthm} guarantees that our previous calculations give us an $+\epsilon k$ approximation to $\HD(T\intervalcc{i}{i + m - 1}, P)$, denote this approximation by $\widehat{\HD}(T\intervalcc{i}{i + m - 1}, P)$. 

    If $k_i = 1$, $P^*, T^*$ have constant run-length encodings and so we can compute their exact $ \HD(T^*\intervalcc{\phi(i)}{\phi(i) + |P^*| - 1}, P^*)$ in constant time, obtaining $\HD(T\intervalcc{i}{i + m - 1}, P)$.

    Otherwise, the index $i$ was discarded for the previous value of $k$, meaning $\HD(T\intervalcc{i}{i + m - 1}, P) >  \min(m, 2\cdot k_i/2)$, so $\HD(T\intervalcc{i}{i + m - 1}, P) >  k_i$. Therefore, $\widehat{\HD}(T\intervalcc{i}{i + m - 1}, P)$ is a multiplicative $(1+\epsilon)$-approximation to $\HD(T\intervalcc{i}{i + m - 1}, P)$.

    Thus, for $n = \Theta(m)$, computing a multiplicative $(1+\epsilon)$-approximation to Text-to-Pattern Hamming Distance problem reduces in $\tilde{O}(n)$ time to $O(\log n)$ instances of computing an additive $+\epsilon k$ approximation to Text-to-Pattern Hamming Distance on strings of length $O(m)$ and run-length encoding $O(k)$.
\end{proof}

We are left to prove \cref{thm:GUthm}.

\begin{proof}[Proof of \cref{thm:GUthm}]
    We follow the proof of \cite{GawrychowskiU18}, using the kernelization technique of Clifford et al. \cite{CliffordFPSS16}. An integer $\pi$ is an $x$-period of a string $S$ of length $m$ if $\HD(S\intervalcc{\pi}{m-1}, S\intervalcc{0}{m - \pi-1})\leq x$.

    \begin{fact}[\cite{GawrychowskiU18, CliffordFPSS16}]\label{fact:period}
        If the minimal $2x$-period of a pattern is $\ell$, then the starting positions of any two occurrences with $\leq x$ mismatches of the pattern are least distance $\ell$ apart.    
    \end{fact}

    The algorithm differentiates between two cases depending on the minimal $O(k)$-period of the pattern $P$. To estimate this period, we use a previous $(1+\epsilon)$-approximation\footnote{Here the $(1+\epsilon)$-approximation guarantees that the returned estimate $\tilde{D}$ satisfies $D\leq \tilde{D}\leq (1+\epsilon)D$} of Karloff \cite{Karloff93}, with $\epsilon = 1$, to match the pattern $P$ against itself in time $O(m \log ^2 m \log |\Sigma|)$. The algorithm can be adapted to work for matching suffixes and prefixes instead of the entire string. This allows us to estimate the period and establish one of two cases: (1) every $4k$-period of $P$ is $\geq k$; (2) there is an $8k$-period of $P$ that is $\leq k$.

    \paragraph{Case 1: No small $4k$-period.} We run Karloff's algorithm again with $\epsilon = 1$ on $T,P$ and identify positions $i$ where the 2-approximation reports $\leq 2k$ mismatches, meaning $\HD(T\intervalcc{i}{i + m - 1},P)\leq 2k$. By \cref{fact:period}, these positions are at least $k$ apart so there are at most $O(m/k)$ of them. We can discard all others as they have $\HD(T\intervalcc{i}{i + m - 1}, P) > k$. For each such $i$, using the technique of Landau and Vishkin \cite{LandauV86}, we count the number of mismatches, either computing $\HD(T\intervalcc{i}{i + m - 1}, P)$ exactly or determining that $\HD(T\intervalcc{i}{i + m - 1}, P)>k$ and discarding it in $O(k)$ time per position. For all positions this take a total of $O(m)$ time.

    \paragraph{Case 2: Small $8k$-period.} Let $\ell\leq k$ be some $8k$-period of $P$. For a string $S$ and $i \in [\ell]$ denote by $\{S\}_{\ell, i}$ the string obtained by taking the characters of $S$ at position $i$ mod $\ell$, $\{S\}_{\ell,i} = S[i]S[i+\ell]S[i+2\ell]\ldots$. We refer to the concatenations of these strings for all indices $i$ as an `$\ell$-encoding' of $S$, $\{S\}_\ell\coloneqq \{S\}_{\ell,0}\{S\}_{\ell,1}\ldots \{S\}_{\ell, \ell-1}$. This representation corresponds to writing the characters of $S$ in a table of width $\ell$ and reading them out column by column.

    Denote by $\runs(S)$ the number of runs in $S$ and denote by $\runs_\ell(S)\coloneqq \sum_{i=0}^{\ell-1} \runs(\{S\}_{\ell,i})$. Note that $\runs_\ell(S) \geq \runs(\{S\}_\ell)$. We will use the following two lemmas from \cite{CliffordFPSS16}.

    \begin{lemma}[\cite{GawrychowskiU18, CliffordFPSS16}]
        If $P$ has an $8k$-period $\ell$ for $\ell\leq k$ then $\runs_\ell(P)\leq 9k$.
    \end{lemma}
    
    \begin{lemma}[\cite{GawrychowskiU18, CliffordFPSS16}]\label{lem:t'}
        Denote by $T_L$ the longest suffix of $T\intervalcc{0}{m-1}$ with $\runs_\ell(T_L)\leq 11k$ and by $T_R$ the longest prefix of $T\intervalcc{m}{2m-1}$ with $\runs_\ell(T_R)\leq 11 k$. Let $T'=T_LT_R$. Then for every $i$ such that $\HD(T\intervalcc{i}{i+m-1},P)\leq k$, $T\intervalcc{i}{i+m-1}$ is contained in $T'$.
    \end{lemma}

    Thus in this case, our problem is reduced to solving Text-to-Pattern Hamming Distance on $T',P$, both of whose $\ell$-encoding has few runs ($\runs_\ell(T')\leq 22k)$. By using $O(\ell) \leq O(k)$ padding of distinct symbols, we can assume $|T'|$ and $|P|$ are both divisible by $\ell$ without changing the complexity of our problem. Denote $|T'| = m_1\cdot \ell, |P|=m_2\cdot \ell$ for some $m_1\geq m_2$.

    We now rearrange $T'$ and $P$ to take advantage of the structure of their $\ell$-encoding. Define 
    \[T^* \coloneqq \{T'\}_\ell \{T'\intervalcc{\ell }{m_1\ell-1}\#^\ell\}_\ell,~~~
    P^* \coloneqq \{P\$^{(m_1-m_2)\ell}\}_\ell.\]

    If we think of the characters of $T$ arranged in a table of width $\ell$, $T^*$ is a string of length $2m_1\ell$ including the columns of the table, followed by each column with the first row entry removed and a $\#$ symbol added at the end. Similarly, $P^*$ is a string of length $m_1\ell$ which includes the columns of the table representation of $P$, with $m_1-m_2$ $\$$-signs added after each one. These string preserve the Hamming Distance of $T',P$ in the following sense.

    \begin{lemma}[\cite{GawrychowskiU18}]\label{lem:maintainHD}
        For any $0\leq \alpha \leq (m_1-m_2)\ell$ write $x = \lfloor \alpha/\ell\rfloor$ and $y = \alpha ~\text{mod}~\ell$. Define $\beta = x + y\cdot m_1$, then
        \[
        \HD(T'\intervalcc{\alpha }{\alpha + m_2\cdot \ell-1}, P) = \HD(T^*\intervalcc{\beta}{\beta + m_1\cdot \ell-1}, P^*) - (m_1-m_2)\cdot \ell.
        \]
    \end{lemma}

    \begin{proof}
        Define the character-inequality function $\delta(a,b) \coloneqq \mathds{1}_{a \neq b}$. We have
        \[
        \HD(T'\intervalcc{\alpha }{\alpha + m_2\cdot \ell-1}, P ) = \sum_{i=0}^{m_2-1}\sum_{j=0}^{\ell-1} \delta(T'[x\ell + y + i\ell j], P[i\ell + j]) = \star.
        \]

        For any $i\in [m_2],j\in [\ell]$ we have $P[i\ell + j] = P^*[i + j\cdot m_1]$. For $ j \in [\ell - y]$ we have $T'[x\ell + y + i\ell + j] = T^*[(x+i) +(y+j)m_1]$ and for $\ell - y \leq j < \ell$ we have $T'[x\ell + y + i\ell + j] = T^*[(x+i) + (y+j-\ell)m_1 + \ell\cdot m_1] = T^*[(x+i) + (y+j)m_1]$. For $i\geq m_2$ and any $j\in [\ell]$, $P^*[i+j\cdot m_1]=\$$, which creates a mismatch with any character in $T^*$. Thus,
        \begin{align*}
            \star &= \sum_{i=0}^{m_2-1}\sum_{j=0}^{\ell-1} \delta(T^*[(x+i) + (y+j)m_1], P^*[i + j\cdot m_1]) \\
            &= \sum_{i=0}^{m_1-1}\sum_{j=0}^{\ell-1} \delta(T^*[(x+i) + (y+j)m_1], P^*[i + j\cdot m_1]) - (m_1-m_2)\ell\\
            &= \HD(T^*\intervalcc{\beta }{\beta + m_1\cdot \ell-1}, P^*) - (m_1-m_2) \ell.
        \end{align*}
    \end{proof}

    We can now complete the proof. If $|T'|<m$, by \cref{lem:t'} we can discard all positions. Otherwise let $t = m - |T_L|$ be the starting position of $T'$. Again by \cref{lem:t'} we can discard any $i \notin \intervalcc{t}{t + (m_1-m_2)\ell}$ as $T\intervalcc{i}{i+m}$ is not contained in $T'$. For the remaining positions $i$ define $\alpha_i = i - t$ and define $\beta_i$ as in \cref{lem:maintainHD}. Taking $\phi(i) \coloneqq \beta_i$  and $c\coloneqq (m_1-m_2)\ell$ gives us the desired guarantees - every $i$ which is not discarded in this case has, 
    \[
    \HD(T\intervalcc{i}{i + m - 1}, P) = \HD(T^*\intervalcc{\phi(i)}{\phi(i) + |P^*| - 1}, P^*)-c.
    \]
    The strings $T^*, P^*$ have length $O(m)$ and RLE $O(k)$, as required.
\end{proof}

\end{document}